\documentclass[onecolumn,prd,amsmath,amssymb,superscriptaddress,nofootinbib]{revtex4-1}
\usepackage{color}
\pdfoutput=1
\usepackage{graphicx}
\usepackage[latin1]{inputenc}
\usepackage{hyperref}
\usepackage{epstopdf}
\usepackage{slashed}
\usepackage{cancel}
\usepackage{epsfig,psfrag,graphics,verbatim}
\usepackage{dcolumn}
\usepackage{bm}
\usepackage{slashed}
\usepackage[]{units}
\usepackage{ulem}
\usepackage{url}
\usepackage{hyperref}
\usepackage{array}
\usepackage{makecell}
\usepackage{soul}

 \def\be   {\begin{equation}}  
  \def\ee   {\end{equation}}
 \def\ba   {\begin{array}}     
  \def\ea   {\end{array}}
 \def\bea  {\begin{eqnarray}}  
  \def\eea  {\end{eqnarray}}
 \def\bean {\begin{eqnarray*}}  
 \def\eean {\end{eqnarray*}}

 \def\lee { \left( }
\def\rii { \right) }

\def\nud {0\nu\beta\beta}

\definecolor{darkgreen}{rgb}{0,0.5,0}

\newcommand{\AddrSDU}{
CP$^{3}$-Origins, University of Southern Denmark, Campusvej 55, DK-5230 Odense M, Denmark
}

\newcommand{\AddrDortmund}{
Fakult\"at f\"ur Physik, Technische Universit\"at Dortmund,
44221 Dortmund, Germany
}

\newcommand{\AddrAU}{
Department of Physics and Astronomy, University of Aarhus, Ny Munkegade 120, DK-8000 Aarhus C, Denmark
}

\begin{document}

{\small
\begin{flushright}
CP3-Origins-2018-003 DNRF90 \\
DO-TH 18/03 
\end{flushright} }

\title{Asymmetric dark matter, baryon asymmetry and lepton number violation}

\author{Mads T. Frandsen}\email{frandsen@cp3.sdu.dk}
 \affiliation{\AddrSDU}
 
 \author{Claudia Hagedorn}\email{hagedorn@cp3.sdu.dk}
 \affiliation{\AddrSDU}
 
\author{Wei-Chih Huang}\email{huang@cp3.sdu.dk}
 \affiliation{\AddrSDU}
 \affiliation{\AddrDortmund}

 \author{Emiliano Molinaro}\email{molinaro@cp3-origins.net}
 \affiliation{\AddrSDU}
 \affiliation{\AddrAU}

\author{Heinrich P\"as}\email{heinrich.paes@tu-dortmund.de}
\affiliation{\AddrDortmund}

\begin{abstract}

We study the effect of lepton number violation (LNV) on baryon asymmetry, generated in the early Universe, in the presence
of a dark sector with a global symmetry $U(1)_X$, featuring asymmetric dark matter (ADM). 
We show that in general LNV, observable at the LHC or in neutrinoless double beta decay experiments, cannot wash out a baryon asymmetry generated at higher scales,
unlike in scenarios without such dark sector. An observation of LNV at the TeV scale may thus support ADM scenarios. Considering several models with
different types of dark matter (DM), we find that the DM mass is of the order of a few GeV or below in our scenario.

\end{abstract}

\maketitle

\section{Introduction}
\label{sec:Introduction}

Both, the cogent astrophysical evidence for DM and the baryon asymmetry of the Universe,
  call for the existence of new physics beyond the Standard Model (SM). 
   Interestingly, the fractions of the critical density $\rho_c$ of the Universe in baryons and DM are
rather close~\cite{Ade:2015xua,Patrignani:2016xqp}: 
\begin{equation}
\Omega_\text{B}\, h^2 = 0.02226 ± 0.00023 \;\; \mbox{and} \;\; \Omega_{\rm DM}\,h^2 = 0.1186 ± 0.0020,  
\label{Omegas}
\end{equation}
corresponding to  
\begin{equation}
\Omega_\text{DM} \simeq 5.3 \, \Omega_\text{B},
\label{omegaratio}
\end{equation}
where $h=0.68$ is the scale factor for the Hubble expansion rate~\cite{Patrignani:2016xqp}. 
 This coincidence might hint at a relation between the production mechanisms of the baryon asymmetry and the observed DM relic abundance.
  Such a relation can be realized in the context of ADM scenarios~\cite{Nussinov:1985xr}, see Refs.~\cite{Zurek:2013wia, Petraki:2013wwa} for recent reviews.
  Within this framework, either DM particles or antiparticles remain in the Universe due to an asymmetry, 
  similar to the one among baryons and antibaryons. To distinguish between particles and antiparticles requires the existence of 
 a conserved global quantum number $X$ in the dark sector. Then, an $X$ asymmetry results in an asymmetry among the number of DM particles and antiparticles. 
  This can be realized, if e.g. the dark sector possesses a global symmetry $U(1)_X$.
  
The explanation of neutrino masses also requires physics beyond the SM. Most of the mechanisms, giving mass to neutrinos, assume that they are Majorana
 particles and thus lepton number ($L$) to be violated by two units~\cite{Yanagida:1979as,Glashow:1979nm,GellMann:1980vs,Mohapatra:1979ia,Magg:1980ut,Schechter:1980gr,Mohapatra:1980yp,Foot:1988aq,Babu:2001ex,deGouvea:2007qla}. Leptogenesis is a prominent mechanism for generating the baryon asymmetry~\cite{Fukugita:1986hr}, see Refs.~\cite{Davidson:2008bu,leptogenesis:A01,leptogenesis:A02,leptogenesis:A03,leptogenesis:A04,leptogenesis:A05,leptogenesis:A06} for recent
 reviews, which is based on LNV processes efficient at high energy scales. A positive signal in
 neutrinoless double beta~($\nud$) decay experiments 
 would be a sign of LNV and prove the Majorana nature of neutrinos, see Refs.~\cite{Bilenky:2014uka,Pas:2015eia,DellOro:2016tmg}
 for recent reviews. As shown in Refs.~\cite{Deppisch:2015yqa,Deppisch:2017ecm}, if $\nud$ decay is due to light neutrino masses, such LNV corresponds  
  to washout effects of the $L$ asymmetry that are efficient only for $T \gtrsim 10^{11}$ GeV. 
 
 As emphasized recently in Refs.~\cite{Deppisch:2013jxa,Deppisch:2015yqa,Deppisch:2017ecm}, any future observation of LNV, such as same-sign dilepton signals at the LHC~\cite{Deppisch:2013jxa} and contributions to $\nud$ decay due to new particles with masses around the TeV scale~\cite{Deppisch:2015yqa,Deppisch:2017ecm},  
will imply LNV processes efficient at temperatures $T \sim 1 \, \rm TeV$. 
These in combination with SM sphaleron processes~\cite{Kuzmin:1985mm,Arnold:1987mh}, that violate $(B + L)$ and are efficient above the sphaleron decoupling temperature $T_{\rm sph}$, can lead to the washout of both baryon number ($B$) and $L$ asymmetries~\cite{Harvey:1990qw}. 
Consequently, a future discovery of LNV processes can falsify  mechanisms of high-scale baryogenesis and leptogenesis. 
Typical models which lead to such LNV processes are radiative neutrino mass models~\cite{Restrepo:2013aga, Cai:2017jrq}.

We therefore investigate the impact of LNV processes, efficient at $T \sim 1\, \rm TeV$, on particle-antiparticle asymmetries in the dark and the visible sectors in ADM scenarios,
 where $B$ and $X$ asymmetries are correlated. We focus on LNV processes that violate $L$ by
 two units. We discuss several models with a symmetry $U(1)_X$ in the dark sector. DM can be a scalar or a fermion, it can be an elementary particle or a composite state
 and there might exist more than one species of DM. We find that in all models it is possible to preserve $B$ and $X$ asymmetries, even in the presence of LNV processes
 at temperatures $T \sim 1 \,\rm TeV$. We can identify two classes of models: those in which the dark sector changes the condition for hypercharge ($U(1)_Y$) neutrality of the Universe~\cite{Harvey:1990qw,Antaramian:1993nt}  and those in which the dark sector also modifies SM sphaleron processes~\cite{Barr:1990ca}.
 The latter implies a transfer of the particle-antiparticle asymmetries between the dark and the visible sectors.  
 Taking into account LNV processes, efficient at $T \sim 1\, \rm TeV$, further constrains the allowed range for the DM mass and requires it to be lower than a few GeV in 
 the models considered, if the ratio of $\Omega_{\text{DM}}$ and $\Omega_{\text{B}}$ in Eq.~(\ref{omegaratio}) should be correctly reproduced.

This paper is organized as follows: In section~\ref{sec:framework},
we list the equations relevant for the different chemical potentials, discuss the features which are necessary
for preventing the washout of the different asymmetries through LNV processes, and then analyze the correlation between the different asymmetries in a model-independent way. 
In particular, we show that LNV processes further constrain the DM mass.
In sections~\ref{sec:example_calc} and \ref{sec:sphalerons} we discuss concrete examples, where either the condition for hypercharge neutrality or also sphaleron processes are modified
through the dark sector, and calculate the correlation between the different asymmetries as well as the upper limit on the DM mass, arising if LNV processes are also considered.
We summarize our findings and conclude in section~\ref{sec:conclusions}.

\section{General Framework}
\label{sec:framework}

The following analysis is carried out for temperatures above the electroweak phase transition~(EWPT).
All statements are valid for situations in which $B$, $L$ and $X$ asymmetries are exclusively generated above  the electroweak~(EW) scale.
The origin of the asymmetries is not specified. They might, for example, be dynamically generated in the early Universe via some unknown new 
physics~\cite{Sakharov:1967dj}. We focus on ADM scenarios, where $\Omega_{\rm DM}$ is only due to the asymmetric component.
This can be justified, if there exists a mechanism, such as the exchange of a light gauge boson or the Sommerfeld enhancement~\cite{ANDP:ANDP19314030302}, 
efficient enough to annihilate the symmetric component.

We first mention the equations relevant for the different chemical potentials and discuss the features, necessary
for preventing the washout of the different asymmetries through LNV processes, efficient at temperatures $T \sim 1\, \rm TeV$. In a second
step, we analyze the correlation between the different asymmetries in a model-independent way and show the additional constraints on the DM mass 
 arising from the presence of LNV processes.

 \subsection{Conditions on Chemical Potentials}
 \label{subsec:chemequi}
 
 Above the EWPT, the SM Yukawa interactions and sphaleron processes are in thermal equilibrium and
the total hypercharge vanishes. 
Thus, the following relations for the chemical potentials $\mu_i$ hold~\cite{Harvey:1990qw}:
 \begin{eqnarray}
 \begin{split}
& -\mu_q + \mu_H + \mu_{d_R}=0 \;\; , \;\; 
-\mu_q - \mu_H + \mu_{u_R}=0 \;\; , \;\;
-\mu_{\ell} + \mu_H + \mu_{e_R}=0 \;\; , \\
& 3 \, (3 \, \mu_q + \mu_{\ell}) =0 \;\; , \;\;
3\, \mu_{q} + 6 \, \mu_{u_R} - 3\, \mu_{d_R} - 3\, \mu_{\ell} -3 \, \mu_{e_R} + 2 \, \mu_H = 0 \;\; . 
\label{eq:SM_chemical}
 \end{split}
 \end{eqnarray} 
 Here, the index $i$ denotes the SM particles: $q$~(left-handed quark doublet), $u_R$~(right-handed $u$-quark), $d_R$~(right-handed $d$-quark), $\ell$~(left-handed lepton doublet), $e_R$~(right-handed charged lepton) and $H$~(Higgs doublet).
The first three conditions originate from the SM Yukawa interactions, while the fourth and fifth conditions correspond to the SM sphaleron processes and hypercharge neutrality of the Universe, respectively. Since we focus on temperatures above the EWPT,
  hypercharge neutrality is required instead of electric charge neutrality, and both components of a left-handed doublet have the same chemical potential.   
For simplicity, we assume that the chemical potentials of the three generations of quarks and leptons are the same, respectively. 
In particular, non-vanishing off-diagonal elements of the quark mixing matrix and efficient EW interactions are responsible for the $B$ asymmetry being equally 
shared by the three quark generations.
In total, there are six chemical potentials and five constraints, leading to one free parameter, which can be chosen as $\mu_{\ell}$.
As a result, both $B$ and $L$ asymmetries can be expressed in terms of $\mu_\ell$~\cite{Kuzmin:1985mm}, see Eq.~(\ref{eq:YBYLSM}).

Once an LNV process, efficient at temperatures $T \sim 1\, \rm TeV$,
is introduced, this leads to an additional condition for the chemical potentials, i.e.,
 a sixth constraint (linearly independent of the ones in Eq.(\ref{eq:SM_chemical})), which implies $\mu_\ell=0$ 
and hence a washout resulting in vanishing $B$ and $L$ asymmetries~\cite{Harvey:1990qw}. 
It is worthwhile to mention that all operators which violate $L$ by two units, that are responsible for e.g. Majorana neutrino masses and $0\nu\beta\beta$ decay,
listed in Refs.~\cite{Babu:2001ex,deGouvea:2007qla}, yield exactly the same constraint:
\begin{align}
\mu_{\ell} +\mu_H =0 \; .
\label{eq:DL=2} 
\end{align}  
This constraint also follows from the Weinberg operator $\left( \ell  H \right)^2$. However, in order to correctly produce the
light neutrino mass scale the coupling of the Weinberg operator is too small to be effective and thus no further condition for the chemical potentials has to be added.
As sphaleron processes are not efficient at converting $B$ and $L$ asymmetries in the SM during or below the EWPT, this washout of the $L$ asymmetry does not imply 
the washout of the $B$ asymmetry.  

As $\nud$ decay only involves electrons in the final state, an operator giving rise to this process is flavor-sensitive, i.e.~it can only lead to the washout of the 
$L$ asymmetry stored in the electron, but not in the muon or tau flavor. However, if lepton flavor is violated at the same time as $L$, the total $L$
asymmetry, stored  in all flavors, will be entirely erased. Throughout this analysis, we assume that also lepton flavor is violated so that the $L$ asymmetry is equally distributed 
among the three lepton generations.

In ADM scenarios, the dark sector possesses a global symmetry $U(1)_X$ and an $X$ asymmetry results in an asymmetry among the number of DM particles and antiparticles.
If at least some particles of the dark sector transform in a non-trivial way under the SM gauge group $SU(3)_C \otimes SU(2)_L \otimes U(1)_Y$, 
a correlation between the asymmetries in the dark and the visible sectors is induced. 
The simplest option to consider is one in which particles of the dark sector only carry hypercharge and a charge under the global symmetry $U(1)_X$. 
Then, only the condition for hypercharge neutrality is modified. The dark and the visible sectors are in contact through Yukawa interactions, see Eq.~(\ref{eq:yuklag}). In this way,
 a transfer between the asymmetries of the dark and visible sectors is possible.
As shown in section~\ref{sec:example_calc}, in the most minimal scenario the dark sector contains two types of particles, the DM particle and the one also charged under $U(1)_Y$. In this case eight chemical potentials have to fulfil six conditions, the first four conditions displayed in Eq.~\eqref{eq:SM_chemical}, the modified condition for hypercharge neutrality, see Eq.~(\ref{eq:chemhypmodhyp}), and the condition due to the interaction connecting the dark and the visible sectors, see Eq.~(\ref{eq:chemyuk}). Thus, two chemical potentials remain as free parameters. If, in addition, $L$ is violated by two units
through processes, efficient at temperatures $T \sim 1\, \rm TeV$, in this scenario, the condition in Eq.~\eqref{eq:DL=2} has to be fulfilled, and thus one of the two free parameters becomes fixed.
 $B$, $L$ and $X$ asymmetries are in turn proportional to the remaining free parameter and, in particular, are in general non-vanishing.

Another example is the model in Ref.~\cite{Barr:1990ca} where fermions $F$ of the dark sector transform as doublets under $SU(2)_L$ and are charged under $U(1)_X$. In this case the SM sphaleron processes are modified and the corresponding condition for the chemical potentials reads:
\begin{align}
3 \, (3 \, \mu_q + \mu_{\ell}) + n_F \mu_F=0 , 
\end{align}
where $n_F$ is the number of generations of $F$ and $\mu_F$ the chemical potential of $F$. As a consequence, an $X$ asymmetry in the dark sector, proportional to $\mu_F$, becomes 
related to $B$ and $L$ asymmetries in the visible sector and the transfer of asymmetries between the dark and the visible sectors is enabled.
 For simplicity, the decoupling temperature of the modified sphaleron processes is assumed to be similar to the one of the SM sphaleron processes.
In general, not only the sphaleron processes are modified in such scenarios, but also the condition, derived from hypercharge neutrality, as particles of the dark sector also carry
hypercharge. As discussed in section~\ref{sec:sphalerons}, the conditions due to Yukawa interactions, sphaleron processes, hypercharge neutrality and interactions involving only
the dark sector leave two of the chemical potentials as free parameters. Thus, including interactions which violate $L$ by two units reduces the number of free parameters to one
and thus correlates $B$, $L$ and $X$ asymmetries, but still allows them to be non-zero. Instead of LNV interactions one can also consider other constraints leading to one additional condition
on the chemical potentials. One example is the preservation of $(B-L)$, as pointed out in Ref.~\cite{Barr:1991qn}. In this scenario a $B$ asymmetry can thus be produced even without the 
violation of $(B-L)$. As is clear, the simultaneous presence of both, LNV interactions, efficient at temperatures $T \sim 1\, \rm TeV$, and the preservation of $(B-L)$, would lead to the vanishing
of all asymmetries, since then all chemical potentials would become fixed.

In cases where the dark sector only contains the DM particle $\chi$ and an interaction between the dark and the visible sectors, including the neutrino~\cite{Kaplan:2009ag,Buckley:2010ui} and neutron~\cite{Farrar:2005zd,Kaplan:2009ag} portal
\begin{align}
\label{portals}
\chi^2 \lee \ell H \rii^2 \,, \quad\quad \chi d_R d_R u_R~(\text{or } \chi^2 d_R d_R u_R) \,, 
\end{align}
 transfers asymmetries between the two sectors, a $B$ asymmetry will be washed out by LNV processes, efficient at temperatures $T \sim 1\, \rm TeV$.  
 Both, the latter and the interaction between the dark and the visible sectors, imply an extra condition for the chemical potentials, while the conditions in Eq.~(\ref{eq:SM_chemical}) are still valid. So, in total seven conditions 
 have to be fulfilled by the seven chemical potentials of SM particles and $\chi$, meaning that these in general have to vanish. As a consequence, not only the asymmetries in the visible
 sector are erased by LNV processes, but also any $X$ asymmetry such that the ADM scenario becomes invalidated as well.
 
We would like to point out that the previous arguments are based on the assumption that all interactions, including the 
 Yukawa interactions, sphaleron processes, interactions between the dark and the visible sectors and LNV processes, are simultaneously in thermal equilibrium.
If, for example, a certain Yukawa coupling is too small to be effective at the time, when sphaleron processes are efficient, or the interactions between the dark and the visible sectors 
decouple before the sphaleron processes become operative, the asymmetries of the two sectors can still be correlated even in the presence of LNV processes.
Examples for these cases can be found in Refs.~\cite{Fornal:2017owa} and \cite{Antaramian:1993nt}. In Ref.~\cite{Fornal:2017owa} a model is considered
with an additional $SU(2)$ gauge symmetry leading to new sphaleron processes. It is essential that some of the Yukawa couplings are small 
 in order not to destroy the $X$ asymmetry. 
  In the model in Ref.~\cite{Antaramian:1993nt} a new charged particle, being part of the dark sector, is added
to the SM. Then, an asymmetry in the dark sector entails a $B$ asymmetry, if the asymmetry in the dark sector is generated after
the interaction of the new charged particle with the visible sector~(needed for it to decay into three charged leptons) becomes ineffective, but above the 
sphaleron decoupling temperature $T_{\rm sph}$.
Before studying the correlation between $B$, $L$ and $X$ asymmetries in concrete ADM scenarios without and with LNV processes, that violate $L$ by two units, 
 in sections~\ref{sec:example_calc} and~\ref{sec:sphalerons}, we discuss the correlation between the different asymmetries in a model-independent way in the following subsection. 
 
 \subsection{Correlation of Asymmetries}
 \label{subsec:corasymm}

In the following, we  consider a theory with the SM and a dark sector, charged under a global symmetry $U(1)_X$.
   $X$ is a new quantum number, assumed to be different from $B$, $L$ and $B-L$. 
In order to be as model-independent as possible, we only require that some of the particles of the dark sector have 
 EW interactions, such that the conditions corresponding to SM sphaleron processes and/or hypercharge neutrality of the Universe, see Eq.~(\ref{eq:SM_chemical}), are modified.
 We further assume that the lightest particle $\chi$ in the dark sector, the DM particle, is neutral and stable. For simplicity, we consider only one species of DM particles in the following.
   Its stability is guaranteed by the global symmetry $U(1)_X$.\footnote{It could be also a discrete subgroup of $U(1)_X$ or another symmetry
 responsible for the stability of $\chi$.}  
  Based on the assumptions made in the preceding subsection, the particles remaining in thermal equilibrium above the sphaleron decoupling temperature, which in the SM is $T_{\rm sph}\approx 135$ GeV \cite{Burnier:2005hp}, are all SM particles and (elementary) scalars and fermions of the dark sector. Their chemical potentials can be expressed in terms of two of them.
  This is equivalent to expressing the different asymmetries as a linear combination of two of them. We choose them in the following as $(B-L)$ and $X$ asymmetry which are
  denoted by $\Delta(B-L)$ and $\Delta X$. We define $B$ and $L$ asymmetries as well as $\Delta (B-L)$ taking into account only SM particles.
 
 The observed baryon asymmetry,  $Y_{\Delta B}=(8.65\pm 0.09)\times 10^{-11}$~\cite{Ade:2015xua,Patrignani:2016xqp},
can be written as
 \begin{equation}
 Y_{\Delta B} \; = \; a_1\, Y_{\Delta(B-L) }\,+\, a_2\, Y_{\Delta X}\,,
 \label{YBAnsatz}
 \end{equation}
 with $a_2\neq 0$ and $a_1\neq 1$.
 The second inequality, $a_1\neq 1$,
 implies that there is no direct proportionality  \textit{ab initio} between the $L$ asymmetry $\Delta L$ and $\Delta X$. 
 Alternatively, one can express $\Delta (B-L)$ as a function of the $B$ asymmetry $\Delta B$ and $\Delta L$
 \begin{equation}
 Y_{\Delta(B-L)}\; = \; \frac{a_2}{1-a_1}\,Y_{\Delta X}\,-\, \frac{1}{1-a_1}\,Y_{\Delta L}\,.
 \label{YBL}
 \end{equation}
 As $\Delta B$ and $\Delta L$ refer only to the SM particles, they
 are defined, in terms of the quark and lepton chemical potentials, as
 \begin{eqnarray}
 \label{eq:YBLmu}
 	Y_{\Delta B} & = & \frac{45}{4\,\pi^2 g_\star}\left( 2\,\mu_q+\mu_{u_R}+\mu_{d_R}\right)\,,\\
	Y_{\Delta L} & = & \frac{45}{4\,\pi^2 g_\star}\left( 2\,\mu_\ell+\mu_{e_R}\right)\,,
 \end{eqnarray}
 where  $g_\star$ as a function of $T$ counts the total number of relativistic degrees of freedom in the thermal plasma.
 If only SM particles are in thermal equilibrium at $T\gtrsim T_{\rm sph}$, we have $a_{2,\text{SM}}=0$ and $g_\star=106.75$. We can derive 
 \begin{equation}
 \label{eq:YBYLSM}
 	Y_{\Delta B}  =  - \frac{15}{\pi^2 g_\star} \mu_\ell \quad\text{and}\quad 	Y_{\Delta L} \; = \; \frac{17}{7} \frac{45}{4\pi^2 g_\star} \mu_\ell \,,
 \end{equation}
from Eq.~(\ref{eq:SM_chemical}) which implies $a_{1,\text{SM}}= 28/79$, as is well-known.

 Similar to Eq.~(\ref{eq:YBLmu}), the yield $Y_{\Delta X}$ of the dark sector is given by the sum of the yields $Y_{\Delta i}$ of all particle species $i$, including the DM particle $\chi$,  carrying a non-zero $U(1)_X$ charge,
   \begin{equation}
   \label{eq:yield-exprX}
	Y_{\Delta i}(z_i) \; = \;  \frac{15\, g_i}{4\,\pi^2\, g_\star}\, \mu_i\,\zeta(z_i)\,,
\end{equation}
where  $g_i$ denotes the internal degrees of freedom of the particle $i$ (e.g., spin and gauge multiplicities), $\mu_i$ is the chemical potential and $z_i= M_i/T$ with $M_i$ being the mass 
of particle $i$ and $T$ the temperature of the plasma.
The function $\zeta (z)$ depends on the particle statistics and is given by
\begin{equation}
	\zeta(z_i) \; = \; \frac{6}{\pi^2}\int_{z_i}^\infty dx \,x\, \sqrt{x^2 - z_i^2} \frac{{\rm e}^x}{\left( {\rm e}^x - \eta_i\right)^2}\,,
\end{equation}
with $\eta_i= 1~(-1)$ for a boson~(fermion). For a relativistic boson~(fermion) with $z_i \ll 1$, we have $\zeta(z_i)\approx 2~(1)$. 
Notice that the relation in Eq.~\eqref{eq:yield-exprX} is valid under the assumption that the number density asymmetry of the particle in the early Universe is much smaller than its equilibrium number density, that is $\mu_i/T \ll 1$. 
For simplicity, we assume that $g_\star$ is always very close to the value of the SM.

 The precise values of the coefficients  $a_1$ and $a_2$ in Eq.~\eqref{YBAnsatz} are model-dependent, see Tab.~\ref{Tab1} and sections~\ref{sec:example_calc} and \ref{sec:sphalerons}.
 They are found by solving the system of conditions for the chemical potentials of all particles in the dark and the visible sectors.
 If an asymmetry $\Delta B$, $\Delta L$ or $\Delta X$ is created, it will be re-shuffled due to Yukawa interactions and sphaleron processes.
 How it is exactly re-distributed among $\Delta B$, $\Delta L$ and $\Delta X$ is encoded in $a_1$ and $a_2$ in Eq.~\eqref{YBAnsatz}. 
 
 Using Eq.~(\ref{YBAnsatz}) one can derive a correlation between the fraction $\Omega_\text{B}$ of $\rho_c$ in baryons and the fraction $\Omega_\chi$ in the DM particle $\chi$
 \begin{equation}
 \label{OmegaBrel}
 \Omega_\text{B}\; = \; a_1\, M_p\, \frac{s\,Y_{\Delta(B-L)}}{\rho_c}\,+\,a_2\,\frac{M_p}{M_\chi}\,\Omega_\chi\,,
 \end{equation}
where $M_p$ and $M_\chi$ denote the proton and DM mass, respectively. Here, $s$ is the entropy density and as numerical values of $M_p$ and $\rho_c$ we use 
$M_p \approx 938.27 \, \rm MeV$ and
$\rho_c=1.88\times 10^{-26} h^2 {\rm kg/m}^3$. 
The fraction $\Omega_\chi$ is defined as $\Omega_\chi\equiv M_\chi (n_\chi-n_{\bar{\chi}})/\rho_c$ with $n_\chi$ and $n_{\bar{\chi}}$ being the number
densities of DM particles and antiparticles, respectively.
 From the phenomenological constraint $\Omega_\chi \leq \Omega_{\rm DM}$, one can derive an upper limit on $M_\chi$:  
 \begin{equation}
 	M_\chi\;\leq\; M_p\,\frac{|a_2|}{\left| 1\,-\,a_1\,\frac{Y_{\Delta(B-L)}}{Y_{\Delta B}}\right|}\,\frac{\Omega_{\rm DM}}{\Omega_{\rm B}}\,.
	\label{DMmasslimit}
 \end{equation}
 If we assume that $\chi$ makes up all DM, equality holds. 
  The asymmetry $Y_{\Delta(B-L)}$ and the coefficients $a_{1,2}$ in Eq.~(\ref{DMmasslimit}) are evaluated at the sphaleron decoupling temperature $T_{\rm sph}$.
As one can see, the requirement of correctly reproducing the observed value of $\Omega_{\rm B}$ constrains both, the DM mass $M_\chi$ and 
the requisite amount of asymmetry $Y_{\Delta(B-L)}$ (or $Y_{\Delta L}$), to be generated by some mechanism.
 In Ref. \cite{Perez:2013tea} a relation similar to the one in Eq.~(\ref{DMmasslimit}) has been obtained in a specific model where a generalized baryon number is gauged,
 see section~\ref{sec:sphalerons} for further details, in particular Eq.~(\ref{eq:doubletmodel1}).

\begin{figure}
\centering
\includegraphics[width=0.48\textwidth]{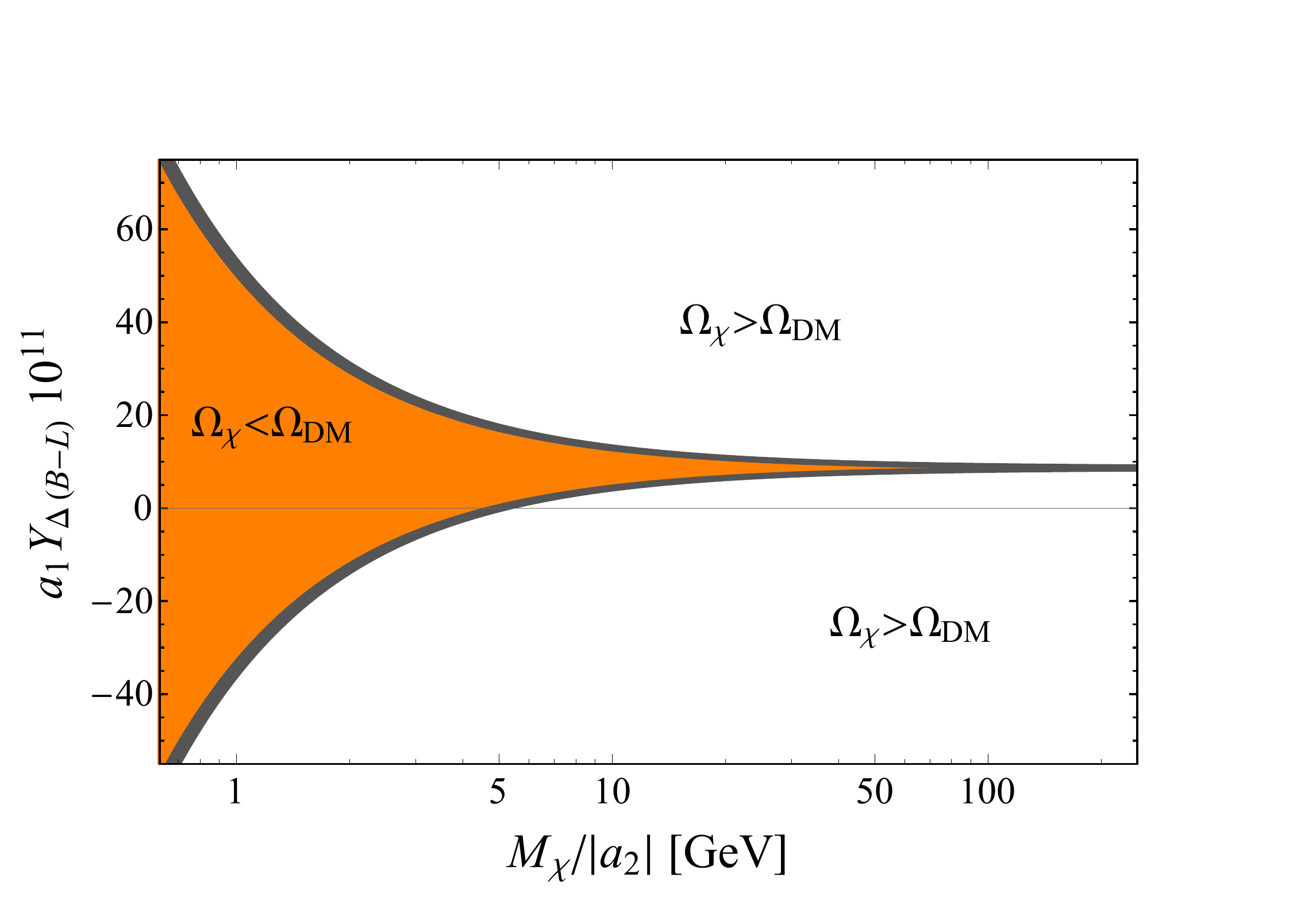}  \includegraphics[width=0.48\textwidth]{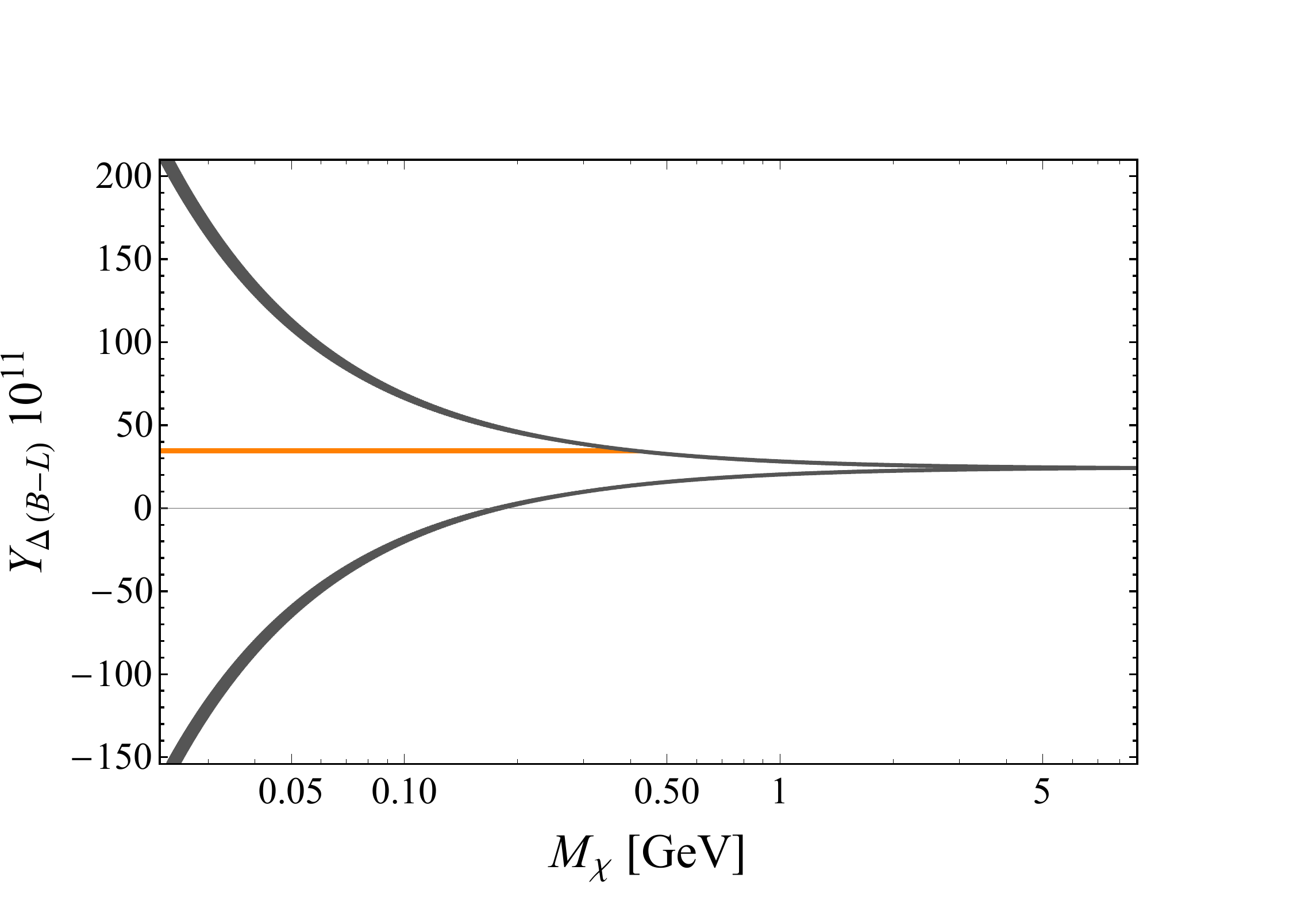}
\caption{
\textbf{Left panel}. The upper limit on the DM mass $M_\chi$, Eq.~(\ref{DMmasslimit}), for arbitrary choices of $a_1$ and $a_2$.
The black thick curves correspond to $ \Omega_{\rm DM}\,h^2 = 0.1186 \pm 0.0040$~($95 \%$ CL).
 In the entire plane $Y_{\Delta B}$ matches the observed value at the $95 \%$ CL. 
 The two regions with $\Omega_\chi>\Omega_{\rm DM}$ are excluded.
\textbf{Right panel}. The allowed region $\Omega_{\chi}\leq \Omega_{\rm DM}$ (thick orange line) in the plane $Y_{\Delta(B-L)}$ vs $M_\chi$, evaluated for one particular model with $a_1=5/14$, $a_2=-1/28$ and $a_3=7$,
see table~\ref{Tab1} and section~\ref{sec:example_calc}, if LNV processes, that violate $L$ by two units, are efficient at temperatures $T \sim 1\, \rm TeV$.}
\label{Fig1}
\end{figure}

 In figure~\ref{Fig1} (left panel) we display, for arbitrary values of  $a_1$ and $|a_2|$, 
 the allowed region, $\Omega_\chi \leq \Omega_\text{DM}$ in orange, while
 the black thick curves indicate the values of $M_\chi$ and $Y_{\Delta(B-L)}$ which saturate the upper limit in Eq.~\eqref{DMmasslimit}.
 For $M_\chi\gtrsim 50 \,|a_2|$ GeV this requires a fine-tuned value of $Y_{\Delta(B-L)}\approx 9\times 10^{-11}/a_1$.
   The thickness of the black curves corresponds to the 95\% CL allowed regions for
$\Omega_\text{B}$ and $\Omega_\text{DM}$, see Eq.~(\ref{Omegas}). The two regions corresponding to $\Omega_\chi>\Omega_{\rm DM}$ are excluded, unless $\Omega_\chi$ is diluted at $T<T_{\rm sph}$ via new entropy injection as demonstrated in, e.g., Ref.~\cite{Bramante:2017obj}.
 Notice that, the observed value of $Y_{\Delta B}$ can be reproduced, even if  $(B-L)$ is conserved in the early Universe, i.e. $Y_{\Delta (B-L)}=0$.
 In this particular case, the asymmetries of DM particles and baryons are directly proportional, as has been realized in Ref.~\cite{Barr:1991qn}.
 
 %
%

Now we assume that LNV processes that violate $L$ by two units are efficient at temperatures $T \sim 1$ TeV.
As argued in section~\ref{subsec:chemequi}, these lead to one additional condition to be fulfilled by the chemical potentials, see Eq.~(\ref{eq:DL=2}),
which reduces the number of free parameters among the chemical potentials to one.
Consequently, the asymmetries $\Delta L$ and $\Delta X$ become proportional to each other
\begin{eqnarray}
  Y_{\Delta L} &= & a_2\, a_3\, Y_{\Delta X}\,, 
  \label{YL2}
\end{eqnarray}	
with $a_3\neq 0$. Hence, from Eqs.~\eqref{YBAnsatz} and \eqref{YBL}, we have
\begin{equation}
	Y_{\Delta B}  =  \frac{a_2\,(1-a_1\,a_3)}{1-a_1}\, Y_{\Delta X} 
	\label{YBL2}\,,
\end{equation}
which in turn implies
 \begin{equation}
 	M_\chi\;\leq\; M_p\,\left| \frac{a_2\,(1-a_1\,a_3)}{1-a_1}\right|\,\frac{\Omega_{\rm DM}}{\Omega_{\rm B}}\,.
	\label{DMmasslimit2}
 \end{equation}
Again the coefficients $a_i$ are computed at $T=T_{\rm sph}$, with $1-a_1 \, a_3 \neq 0$. 
Applying  Eqs.~\eqref{YL2} and \eqref{YBL2} to the case of the SM, we have $a_{2, \rm SM}=0$ such that the vanishing of $B$ and $L$ asymmetries follows,
excluding high-scale baryogenesis and leptogenesis, in accordance with the observations made in Refs.~\cite{Deppisch:2015yqa,Deppisch:2017ecm,Deppisch:2013jxa}.

In figure~\ref{Fig1} (right panel) the allowed region of $\Omega_\chi \leq \Omega_\text{DM}$ is shown for a specific choice of $a_1$ and $a_2$, $a_1=5/14$ and $a_2=-1/28$, see table~\ref{Tab1}.
 These values arise in an example, discussed in section~\ref{sec:example_calc}, where the dark sector consists of one complex scalar $S$, being the DM particle,
  and one charged Dirac fermion. If LNV processes, that violate $L$ by two units and that are efficient at temperatures $T \sim 1\, \rm TeV$, are present, $a_3=7$
and the horizontal orange line in the figure indicates the allowed interval of $M_\chi$ and the value of $Y_{\Delta (B-L)}$. In particular, the DM mass is confined to the sub-GeV range, 
$M_\chi\leq 0.44$ GeV. Low values of $M_\chi$ not larger than a few GeV are found in all the discussed ADM scenarios, see table~\ref{Tab1}.
The latter are discussed, beginning with examples where only the condition for hypercharge neutriality is modified in section~\ref{sec:example_calc}
 and followed by models with elementary and composite particles in section~\ref{sec:sphalerons} in which also the sphaleron processes are modified.

 \begin{table}[t!]
\centering
\catcode`?=\active \def?{\hphantom{0}}
\begin{tabular}{!{\vrule width 1pt}@{\quad}>{\rule[-2mm]{0pt}{6mm}}l@{\quad\;}!{\vrule width 2pt}@{\quad\;}c@{\quad}|@{\quad\;}c@{\quad\;}|@{\quad\;}c@{\quad\;}||@{\quad\;}c@{\quad\;}!{\vrule width 1pt}}
 \Xhline{3\arrayrulewidth}
  \rule[0.15in]{0cm}{0cm}{Model}  &   $a_1$ & $a_2$ & $a_3$  & $M_\chi$ [GeV] \\[0.2mm]
 \hline
SM + 1 generation $\{S,F \}$ with charge $Q_X=+1$ & $\frac{5}{14}$ & $\mp\frac{1}{28}$ & $7$ & 0.44 \\[0.2mm] 
  \hline
Ref.~\cite{Barr:1991qn} & $ \frac{32}{83} $ & $-\frac{72}{415}$ & $\frac{83}{270}$ & 1.3\\[0.2mm]
 \hline
Ref.~\cite{Perez:2013tea} & $\frac{32}{99}$ & $\frac{15-14 \, \mathcal{B}_2}{198}$ & $ \frac{ 297 (17 + 2 \, \mathcal{B}_2) }{119 (-15 + 14 \, \mathcal{B}_2)}$  & 0.49 $\left| 2.3 - \mathcal{B}_2 \right|$  \\[0.2mm]
\hline 
Ref.~\cite{Barr:1990ca} with elementary Higgs $H$& $\frac{12}{35}$ & $ - \frac{32}{35}$ & $ \frac{35}{104}$ & 6.5 \\[0.2mm]
\hline
Ref.~\cite{Ryttov:2008xe} with elementary Higgs $H$& $\frac{8}{23}$ & $-\frac{21}{46}$ & $\frac{23}{63}$ & 3.2 \\[0.2mm]
\Xhline{2\arrayrulewidth}
\end{tabular}
\caption{Prediction of $a_1$, $a_2$ and $a_3$ in the discussed models. The mass $M_\chi$
denotes the upper limit on the mass of the DM particle required to correctly reproduce $\Omega_{\rm DM}$, if LNV processes, that violate $L$ by two units, are present. 
The minus~(plus) sign in $a_2$ for the SM extended by one additional generation of a complex scalar $S$ and a Dirac fermion $F$
refers to $S$~($F$) being the DM particle, given $Q_X=+1$. The parameter $\mathcal{B}_2$ is the charge of certain dark sector particles under a generalized baryon number
and is a free parameter.
 Both composite models have been endowed with an elementary Higgs $H$. For details, see sections~\ref{sec:example_calc} and \ref{sec:sphalerons}. 
  \label{Tab1}}
\end{table}

\section{Modification of Hypercharge Neutrality Condition}
\label{sec:example_calc}

In this section, we present and discuss a simple example of the case where only the condition for hypercharge neutrality is modified, while 
 sphaleron processes are unchanged with respect to the SM. 
For this purpose, we consider an extension of the SM with $n_F$ vector-like fermions $F_j$, and $n_S$ complex scalars $S_k$. 
The fermions and scalars mix among each other via mass terms. This mixing leads to  common chemical potentials for fermions and scalars, 
$\mu_F$ and $\mu_S$, respectively.\footnote{Since $F_j$ are vector-like fermions, left-handed and right-handed components have the same chemical potential.}
  For simplicity, all new particle masses are assumed to be much smaller than the sphaleron decoupling temperature,
 i.e. $M_{F_j}, m_{S_k}\ll T_{\rm sph}$. In addition, all particles of the dark sector transform under a global symmetry $U(1)_X$, which guarantees the stability of the lightest one.
 The lightest particle is neutral and plays the role of the DM candidate. It can be either a fermion or a scalar.
    All new particles are neutral under $SU(3)_C\otimes SU(2)_L$ and carry the same $U(1)_X$ charge, up to a sign, which we normalize to $Q_X=+1$ so that $X(F_j)=-X(S_k)=1$.  Either $F_j$ or $S_k$ (but not both) is charged under $U(1)_Y$ such that a coupling to
  right-handed charged leptons $e_{R \alpha}$ is allowed at tree-level
\begin{equation}
\label{eq:yuklag}
\lambda^\alpha_{j k}\, \, \overline{F}_{L j}\, S_k^\ast \, e_{R \alpha} \, .
\end{equation}
Here, for each lepton flavor $\alpha=e, \mu,\tau$, $\lambda^\alpha$ is a complex $n_F\times n_S$ matrix.
If $\lambda^\alpha_{j k}$ is large enough for the Yukawa interactions to be in thermal equilibrium,
this yields a  
new constraint on the chemical potentials, namely
\begin{equation}
\label{eq:chemyuk}
-\mu_F - \mu_S + \mu_{e_R}=0 \, .
\end{equation}
The condition for hypercharge neutrality is modified, since one species of particles of the dark sector is charged under $U(1)_Y$. We thus have
\begin{equation}
\label{eq:chemhypmodhyp}
 3 \, \mu_{q} + 6 \, \mu_{u_R} -  3 \, \mu_{d_R} - 3 \, \mu_{\ell} -3 \, \mu_{e_R} + 2 \, \mu_H - 2 \, n_{F \, (S)} \, \mu_{F \, (S)}=0 \, ,
\end{equation}
in case the fermions (scalars) carry the $U(1)_Y$ charge. 
The asymmetry $Y_{\Delta X}$ is a function of $\mu_F$ and $\mu_S$, see Eq.~(\ref{eq:yield-exprX}),
\begin{equation}
 Y_{\Delta X} \; = \;  \frac{15}{2\,\pi^2\, g_\star}\, \left(n_F \, \mu_F -  n_S\,\mu_S\right)\,.
\end{equation}
Since none of the new particles transforms non-trivially under $SU(2)_L$, sphaleron processes, see
Eq.~(\ref{eq:SM_chemical}), are not altered. 
Taking into account all conditions fulfilled in thermal equilibrium, two chemical potentials remain as free parameters, and we find for the coefficients $a_1$ and $a_2$
\begin{equation}
	a_1  = \; 4\; \frac{ 7 \left(n_S+ n_F \right) + n_S \,n_F }{79 \left(n_S+ n_F \right) +10 \,n_S\, n_F} \;\; \mbox{and} \;\;
	a_2  = \; \mp \, \frac{6 \, n_{F (S)}}{79 \left(n_S+ n_F \right) +10\, n_S \,n_F} \, .
	\label{eq:coeff12hyp}
\end{equation}
Here, the positive sign and the index $S$ in $a_2$ correspond to the case with neutral fermions, while the negative one and the index $F$ refer to the case of neutral scalars.
For the special case of $n_F=n_S=1$, we obtain $a_1=\frac{5}{14}$ and $a_2=\mp \frac{1}{28}$, as reported in table~\ref{Tab1} and used in figure~\ref{Fig1} (right panel).

When introducing LNV processes which violate $L$ by two units, the additional constraint in Eq.~\eqref{eq:DL=2} has to be included. 
Consequently, we obtain a correlation between
$Y_{\Delta L}$ and $Y_{\Delta X}$, as in Eq.~(\ref{YL2}), with the coefficient $a_3$ being
\begin{equation}
a_3 = \; \frac{79 \left(n_S+ n_F \right) +10 \,n_S\, n_F}{ 11 \left(n_S+ n_F \right) +2\, n_S\, n_F}\,.
\label{eq:coeff3hyp}
\end{equation}
For instance, if $n_F=n_S=1$, one has $a_3=7$ which is used in table~\ref{Tab1} and figure~\ref{Fig1} (right panel).
According to Eq.~(\ref{DMmasslimit2}),
we can set an upper limit on the DM mass $M_\chi$:
\begin{equation}
M_\chi \lesssim \left( 10 \; \frac{n_{F \, (S)}}{11 \left(n_S+ n_F \right) +2\, n_S\, n_F} \right) \, \rm GeV ,
\end{equation}
for neutral scalars (fermions). For $n_F=n_S=1$, we find $M_\chi \lesssim 0.44 \, \rm GeV$, as reported in table~\ref{Tab1},
and the requisite amount of asymmetry $Y_{\Delta (B-L)}$ is displayed in figure~\ref{Fig1} (right panel).

One can straightforwardly generalize these computations to the case of massive fermions $F_j$ and scalars $S_k$ and to cases where $Q_X \neq+1$.  
  The asymmetry $Y_{\Delta X}$, see Eq.~(\ref{eq:yield-exprX}), is then given by
 \begin{equation}
	Y_{\Delta X} \; = \;  \frac{15}{4\,\pi^2\, g_\star}\, Q_X \left[ 2\, \mu_F\,\sum\limits_{j=1}^{n_F} \zeta\left(\frac{M_{F_j}}{T_{\rm sph}}\right) \, - \, \mu_S\, \sum\limits_{k=1}^{n_S} \zeta\left(\frac{m_{S_k}}{T_{\rm sph}}\right) \right]\, 
\end{equation}
 and also the coefficients $a_i$ encode this information, e.g.
 \begin{equation}
 	a_1 = \; 4\; \frac{7 \, \left(\sum\limits_{k=1}^{n_S} \zeta\left(\frac{m_{S_k}}{T_{\rm sph}}\right) + 2 \, \sum\limits_{j=1}^{n_F} \zeta\left(\frac{M_{F_j}}{T_{\rm sph}}\right) \right)+
	\left( \sum\limits_{k=1}^{n_S} \zeta\left(\frac{m_{S_k}}{T_{\rm sph}}\right) \right) \, \left( \sum\limits_{j=1}^{n_F} \zeta\left(\frac{M_{F_j}}{T_{\rm sph}}\right) \right)}{79\, \left( \sum\limits_{k=1}^{n_S} \zeta\left(\frac{m_{S_k}}{T_{\rm sph}}\right)+2 \, \sum\limits_{j=1}^{n_F} \zeta\left(\frac{M_{F_j}}{T_{\rm sph}}\right) \right)
			+ 10 \, \left( \sum\limits_{k=1}^{n_S} \zeta\left(\frac{m_{S_k}}{T_{\rm sph}} \right)\right) \, \left( \sum\limits_{j=1}^{n_F} \zeta\left(\frac{M_{F_j}}{T_{\rm sph}}\right) \right)}
 \end{equation}
for the scalars being neutral.
Furthermore, one can imagine a more complicated setup, where each pair of a fermion and scalar
carries a different $U(1)_X$ charge, and either the fermion or scalar of each pair can be neutral. 
In this setup, one may thus realize multi-component DM~\cite{Zurek:2008qg,Batell:2010bp,Aoki:2012ub,Profumo:2009tb,Arcadi:2016kmk,Herrero-Garcia:2017vrl}.

\section{Modification of SM sphaleron processes}
\label{sec:sphalerons}

In this section, we first discuss two models with elementary and then two with composite particles in the dark sector.
In all these models, the SM sphaleron processes are modified by the dark sector.

Models with elementary particles in the dark sector are discussed in Refs.~\cite{Barr:1991qn,Perez:2013tea,Frandsen:2016bke}.
The relevant symmetry group of these models is
 \begin{equation}
SU(2)_L \otimes U(1)_Y\otimes U(1)_\mathcal{B}\otimes U(1)_\mathcal{L}\otimes  U(1)_X \, ,
 \end{equation}
with the last three $U(1)$ symmetries referring to generalized baryon number ($\mathcal{B}$), generalized lepton number ($\mathcal{L}$) and $X$. 
The generalizations $\mathcal{B}$ and $\mathcal{L}$ are introduced, since in the discussed models the particles of the dark sector either also carry
baryon number, as in Ref.~\cite{Perez:2013tea}, or also lepton number, see Refs.~\cite{Barr:1991qn,Frandsen:2016bke}, and we wish to distinguish these 
from $B$ and $L$, only taking into account SM particles.
The dark sector contains the following fermions
 \begin{gather}
\begin{aligned}
\Psi_L &\sim  ( {\bf 2}, -\frac{1}{2},\mathcal{B}_1,\mathcal{L}_1,X_1),\\
\eta_R   &\sim  ( {\bf 1}, -1,\mathcal{B}_1, \mathcal{L}_1,X_1),\\
\chi_R &\sim  ({\bf 1}, 0 ,\mathcal{B}_1 ,\mathcal{L}_1, X_1)\,, \
\end{aligned}\quad\quad
\begin{aligned}
\Psi_R &\sim ({\bf 2}, -\frac{1}{2},\mathcal{B}_2,\mathcal{L}_2,-X_2), \\
\eta_L &\sim  ( {\bf 1},-1,\mathcal{B}_2,\mathcal{L}_2,-X_2), \\ 
\chi_L &\sim ( {\bf 1}, 0, \mathcal{B}_2,\mathcal{L}_2, -X_2) \, ,
\label{eq:doubletmodel1}
\end{aligned}
\end{gather} 
following (a generalization) of the conventions of Ref.~\cite{Perez:2013tea}. 

In Ref.~\cite{Barr:1991qn} the particles of the dark sector carry generalized lepton number and $U(1)_X$ charge.
The charge assignments are $\mathcal{B}_1=\mathcal{B}_2=0, \mathcal{L}_1=\mathcal{L}_2=1, X_1=X_2=1$. 
In addition, three left-handed fermions, being gauge singlets under the SM, with $U(1)_X$ charge 
\begin{align}
\gamma_{1L}   \sim  ( {\bf 1}, 0,0,0,-1) ,  \quad 
\gamma_{2L}  \sim  ({\bf 1}, 0, 0,0,1) ,  \quad 
\gamma_{3L}  \sim  ({\bf 1}, 0, 0,0,0) 
\label{eq:doubletmodel2}
\end{align}
are introduced which have Yukawa interactions with $\Psi_{L,R}$ and the lepton doublets $\ell$ (and the scalar $H^\prime$).
The global symmetry $U(1)_X$ is violated by sphaleron processes, see Eq.~(\ref{eq:modsphele}). The dark sector also contains two scalars, $H^\prime$ and $h$,
which transform as 
\begin{eqnarray}
H' \sim  ( {\bf 2}, \frac{1}{2}, 0,-\mathcal{L}_1,0) 
\, , \quad \quad 
h \sim({\bf 1},0, 0,0,0)  \,.
\end{eqnarray}
The DM particle is the fermion $\gamma_{1L}$, the lightest particle carrying non-zero $U(1)_X$ charge.

In the model in Ref.~\cite{Perez:2013tea} the particles of the dark sector carry generalized baryon number. The symmetry $U(1)_{\mathcal{B}}$ is gauged in this model.
   The corresponding charge assignments are $\mathcal{B}_1-\mathcal{B}_2=-3$, required by gauge anomaly freedom of $U(1)_{\mathcal{B}}$, with $\mathcal{B}_2$ being a free parameter, $\mathcal{L}_1=\mathcal{L}_2=0$ and $X_1=-X_2=1$ under $U(1)_X$. The latter is an accidental global symmetry of the dark sector and its quantum number is denoted with $\eta$ in Ref.~\cite{Perez:2013tea}.
   Note that this symmetry is not violated by sphaleron processes, see Eq.~(\ref{eq:modsphele}). In addition to the fermions, shown in Eq.~(\ref{eq:doubletmodel1}), the dark sector possesses a scalar $S_{\mathcal{B}}$
   which also carries generalized baryon number, $S_\mathcal{B} \sim({\bf 1},0,-3,0,0) $. When $S_{\mathcal{B}}$ acquires a vacuum expectation value, the gauge symmetry $U(1)_{\mathcal{B}}$  breaks spontaneously and contributions to the masses of the particles of the dark sector are generated.
  In the special case, discussed in Ref.~\cite{Perez:2013tea}, the DM particle is the Dirac fermion $\chi=\chi_L+\chi_R$ which is a singlet under the SM gauge group.
 
In both models, the sphaleron processes are modified and the corresponding condition reads 
\begin{align}
\label{eq:modsphele}
3 \, (3 \, \mu_q + \mu_{\ell}) +  \mu_{\Psi_L} - \mu_{\Psi_R}=0 \, . 
\end{align}
In addition, the Yukawa interactions involving particles of the dark sector also lead to conditions on the chemical potentials.
These are, in particular, necessary in order to convert the $X$ asymmetry to an particle-antiparticle asymmetry of the DM particle.

We report in table~\ref{Tab1} the results from these two models for the coefficients $a_i$ as well as the upper limit on the DM mass in the case in which LNV processes, that violate $L$ by two
 units and that are efficient at temperatures $T \sim 1$ TeV, are present.

\vspace{0.2in}

The first ADM model in which sphaleron processes are modified by the dark sector and transfer asymmetries between the dark and the visible sectors has 
been proposed in Ref.~\cite{Barr:1990ca}. 
This model belongs to a class of models where the dark sector contains composite particles and possesses
 a new strongly interacting Technicolor (TC) gauge group $G_{\rm TC}$. 
 The relevant symmetry group of the model is
 \begin{equation}
G_{\rm TC} \otimes SU(2)_L \otimes U(1)_Y\otimes  U(1)_X \, .
 \end{equation}
A minimal set of new fermions, which does not lead to gauge anomalies, is
\begin{equation} 
Q_L=\left(\begin{array}{c} U\\D \end{array}\right)_L \sim ({\bm R_{\rm TC}},{\bf 2},0,x)
, \qquad U_R \sim ( {\bm R_{\rm TC}},{\bf 1},\frac{1}{2},x)
\
, \quad D_R  \sim ({\bm R_{\rm TC}},{\bf 1},-\frac{1}{2},x) \,,
\label{eq:composite}
\end{equation}
where ${\bm R_{\rm TC}}$ is the particle's representation under $G_{\rm TC}$ and $x$ is the charge under the global symmetry $U(1)_X$, 
 analogous to baryon number in the SM. As opposed to the models with elementary particles discussed above, the dark sector does not contain 
  right-handed $SU(2)_L$ doublets, because these models have been devised to break EW symmetry dynamically via condensation. 

The global symmetry $U(1)_X$ is violated by sphaleron processes, see Eq.~(\ref{eq:modsphcomp}), and, for certain choices of the gauge group $G_{\rm TC}$ and the representation ${\bm R_{\rm TC}}$, 
the lightest composite particle in the spectrum is neutral and stable below the sphaleron decoupling temperature~\cite{Chivukula:1989qb,Barr:1990ca,Gudnason:2006yj,Ryttov:2008xe,Frandsen:2009mi}. 
For example, if $G_{\rm TC}=SU(4)$ and ${\bm R_{\rm TC}}$ is the four-dimensional fundamental representation, the state $\chi \sim U D U D$, a singlet under the SM gauge group, is neutral and is the lightest state carrying $U(1)_X$ charge, as discussed in Ref.~\cite{Barr:1990ca}. 
If $G_{\rm TC}=SU(2)$ and ${\bm R_{\rm TC}}$ is the two-dimensional fundamental representation, then $\chi \sim U D$ is neutral, a singlet under the SM gauge group and the lightest state carrying   $U(1)_X$ charge. In this case it is also a Goldstone boson of the strong dynamics, see Ref.~\cite{Ryttov:2008xe}. 
In both models, sphaleron processes are modified and the corresponding condition for the chemical potentials reads
\begin{align}
\label{eq:modsphcomp}
3 \, (3 \, \mu_q + \mu_{\ell}) + {\rm dim}({\bm R_{\rm TC}}) \, \mu_{Q_L} =0 \,. 
\end{align}

This class of models is nowadays constrained by the observation of the Higgs particle. 
While the condensate $\langle U_L \overline{U}_R + D_L\overline{D}_R + {\rm h.c.}\rangle \sim f^3$, with $f$ being the Goldstone boson decay constant, can yield correct EW symmetry breaking, it may not yield a light Higgs-like excitation. A viable variation may arise from coupling  an elementary Higgs $H$ to the dark sector and to the SM fermions via Yukawa interactions~\cite{'tHooft:1979bh,Simmons:1988fu,Dine:1990jd,Carone:2012cd,Galloway:2016fuo,Agugliaro:2016clv,Alanne:2017rrs}.
Hence, in the simplest case the EW scale is set by 
\begin{equation}
v_{\rm EW}^2 = v^2 +  N_D\, f^2  \,,
\end{equation}
where $v$ is the vacuum expectation value of $H$ 
and $N_D$ is the total number of $SU(2)_L$ doublets in the dark sector divided by ${\rm dim}(\bm R_{\rm TC})$, meaning $N_D=1$ in the two examples.  
The Yukawa interactions of the new fermions and $H$ are of the form 
\begin{equation}
\lambda_D \, \overline{Q}_L H \, D_R+ \lambda_U \, \overline{Q}_L \widetilde{H} \, U_R \, .
\end{equation}
Being in thermal equilibrium above the sphaleron decoupling temperature, they provide conditions for the chemical potentials
 equivalent to those, arising from the SM Yukawa interactions.
We refer to Refs.~\cite{Carone:2012cd} and \cite{Galloway:2016fuo,Agugliaro:2016clv,Alanne:2017rrs} for explicit constructions of models with the gauge group $G_{\rm TC}=SU(4)$ and $G_{\rm TC}=SU(2)$ and an elementary Higgs $H$, and to Ref.~\cite{Alanne:2013dra} for how to raise the scale $f$ in such models.

When computing the coefficients $a_i$ for the models, defined in Refs.~\cite{Barr:1990ca,Ryttov:2008xe}, extended with an elementary Higgs $H$,
we do not consider right-handed neutrinos and adapt as normalization of the global quantum number $X$ the one used in Ref.~\cite{Barr:1990ca}.\footnote{The $U (1)_X$ charge of $Q_L$, $U_R$
and $D_R$ is given by one divided by the number of constituents of the DM candidate $\chi$, i.e.~$x=1/4$ and $x=1/2$ for the model in Ref.~\cite{Barr:1990ca} and \cite{Ryttov:2008xe}, respectively.
This definition deviates from the one chosen in Ref.~\cite{Ryttov:2008xe} by a factor $1/\sqrt{2}$, see definition of the generator of $U(1)_{TB}$, corresponding to $U(1)_X$, in 
 Ref.~\cite{Ryttov:2008xe}.} The coefficients $a_i$ can also be found in table~\ref{Tab1}.

\section{Conclusions}
\label{sec:conclusions}

We have considered ADM scenarios with a dark sector which possesses a global symmetry $U(1)_X$ and the impact of LNV processes on the particle-antiparticle
asymmetries in the dark and the visible sectors. We have shown in a model-independent way and with several examples that $B$, $L$ and $X$ asymmetries can
 be preserved in an ADM scenario, even in the presence of LNV processes that violate $L$ by two units and that are efficient at temperatures $T \sim 1$ TeV.
  This is in contrast to models without ADM, where e.g. the observation of LNV in form of same-sign dilepton signals at the LHC can falsify high-scale baryogenesis and leptogenesis
 as has been shown recently in the literature.

 The crucial feature of all scenarios are particles in the dark sector that carry hypercharge and/or transform in a non-trivial way under $SU(2)_L$. In the former case,
 the condition for hypercharge neutrality is modified, while in the second one  sphaleron processes are also altered by the dark sector. In both cases it turns out
 that fulfilling all conditions, derived from Yukawa interactions, hypercharge neutrality and sphaleron processes, leaves two chemical potentials as free parameters. Thus,
 adding one further condition, imposed by LNV processes on the chemical potentials, reduces the number of free parameters to one. Consequently, all asymmetries 
 become correlated, but are still non-zero in general. The conversion of the asymmetries of the dark and the visible sectors is in the presented examples either achieved through Yukawa interactions
 between the two sectors and/or modified sphaleron processes. Including LNV processes, considerably reduces the allowed parameter space of the ADM scenarios. In particular,
 it puts an upper limit on the DM mass of a few GeV. This makes it challenging to test these ADM scenarios with direct DM detection experiments. Therefore, an observation
 of LNV at the LHC or $\nud$ decay can provide a way to probe the considered models and to indirectly also obtain information on the dark sector.

We would like to emphasize that our analysis assumes all interactions, including Yukawa ones, sphaleron processes and 
other asymmetry transfer interactions, to be in thermal equilibrium. Our conclusions thus do not apply to situations where the $B$ asymmetry is generated below 
(or during) the EWPT or the dark sector decouples from the visible one before LNV becomes efficient.

As the discussed LNV processes naturally arise in radiative neutrino mass models and several of these also possess a global symmetry
and feature a viable DM candidate, a detailed study of these models along the lines of the present analysis
would be very interesting.

\section*{Acknowledgments}

MTF and WCH acknowledge partial funding from the Independent Research Fund Denmark, grant number 
DFF 6108-00623. The CP3-Origins centre is partially funded by the Danish National Research Foundation, grant number DNRF90.

\bibliography{0vbb_ADM_v3}
\bibliographystyle{hunsrt}

\end{document}